\documentclass[preprint]{revtex4}
\usepackage{amsmath,amssymb,bm,graphicx,subfigure,enumerate,latexsym,comment,dcolumn}
\usepackage{color}

\begin{document}

\title{Solidification of the Lennard-Jones fluid near the wall in thermohydrodynamic lubrication}

\author{Kouki Nakamura}    

\author{Ryo Ookawa}    

\author{Shugo Yasuda
  \footnote{Electronic mail: yasuda@sim.u-hyogo.ac.jp}}

\affiliation{Graduate School of Simulation Studies, University of Hyogo, Kobe 650-0047, Japan}

\begin{abstract}
	We investigate the thermohydrodynamic lubrication of the Lennard-Jones (LJ) fluid in the parallel-plate channel composed of the LJ particles by using molecular dynamics (MD) simulation.
	We discover a counterintuitive solidification of the LJ fluid near the wall, i.e., \textit{viscous heating-induced solidification}, where solidification occurs only when the viscous heating of the LJ fluid is sufficiently large.
	The solidification mechanism is investigated from both macroscopic and microscopic points of view.
	It is found that the LJ molecules are densely confined in the vicinity of the wall via the thermohydrodynamic transport of the bulk fluid and that when the local density in the vicinity of the wall is close to the solidification line in the phase diagram, the LJ molecules are solidified due to the interaction with the crystallized wall molecules.
	Band formation is also observed in the highly confined regime when the channel width is sufficiently large.
\end{abstract}  

  \maketitle
\section{Introduction}
Lubrication in high-speed mechanical systems may involve complicated phenomena such as thermorheological coupling in the bulk fluid and chemical reactions and phase transitions at the interface.
Computer simulation of such a complicated lubrication system is challenging and important from both engineering and scientific points of view.

For hydrodynamic lubrication problems, the computational fluid dynamics (CFD) approach is usually utilized by employing any constitutive relations, which stem from the microscopic dynamics of the molecules.
Since constitutive relations are usually unknown for complex fluids involved in, e.g., thermorheological coupling, chemical reactions, and phase transitions, it is difficult to apply CFD simulation to complicated lubrication problems ab initio.
Molecular dynamics (MD) simulation is applicable even for complex fluids because the complicated transport phenomena are autonomously reproduced once their molecular models are appropriately specified.\cite{book:89AT,book:08EM,book:05KBA}

In this study, we analyze the high-speed lubrication of the Lennard-Jones (LJ) fluid in relation to thermohydrodynamic coupling and phase transitions.
Instead of conventional fluid simulations, we perform MD simulation of thermal flows in a channel made of molecularly constituted thermal walls. 
Thus, thermohydrodynamic coupling, e.g., \textit{viscous heating}, and phase transition are autonomously reproduced, as naturally occurs in real high-speed lubrication systems.

It is well known that simple liquids may be solidified in molecularly thin layers due to the confinement, i.e., {\em confinement-induced phase transition} \cite{art:95KK,art:98KK,art:06Aetal}.
Recently, we reported in a proceedings paper that a counterintuitive phase transition, i.e., {\em viscous heating-induced solidification}, occurs at the interface in the high-speed lubrication of the LJ fluid, where the channel width is much larger than the molecular size. \cite{art:18YO}.
In this paper, we comprehensively carry out MD simulations with changing channel width, wall speed, wall temperature, fluid density, and wall structure, and aim to unveil the mechanism of the solidification from both microscopic and macroscopic points of view.

MD simulations of thermohydrodynamic lubrication in nanochannels were previously reported in the literatures~\cite{art:97KPY,art:06HO,art:10KBC,art:17GB}.
These studies clarified the distinctive features of nanoscale flows, such as the slip on the boundaries.
In this study, we consider a channel that is much larger than the molecular size, at which macroscopic transport is significant, and focus on the thermohydrodynamic coupling in the bulk fluid and the phase transition at the boundaries.
The target of our study is more relevant to micromechanical engineering involving high-speed mechanical systems.

Incidentally, the multiscale hybrid method of MD and CFD is currently a very active research field \cite{art:03EE,art:03KGHKRT,art:05RE,art:09KS,art:13BLR,art:16ZRE,art:08YY,art:09YY,art:10YY,art:11YY,art:13MYTY}.
The thermohydrodynamic lubrication of polymeric fluids was also investigated by the synchronized MD method, in which a transitional behavior of the polymer conformation due to thermohydrodynamic coupling was clarified \cite{art:14YY,art:16YY,art:19Y}.
Although the multiscale approach is powerful and promising for complex fluids, to further develop the multiscale method, we require more first-principles results obtained through MD simulations.
This study also aims to contribute to this objective.

In the following text, we describe the problem and simulation model considered in this study in Sec.~\ref{section2}.
The simulation results are given in Sec.~\ref{sec3}, where the counterintuitive solidification is revealed.
In Sec.~\ref{sec4}, the solidification mechanism is discussed from both macroscopic and microscopic points of view.
Finally, we give the concluding remarks in Sec.~\ref{sec5}.

\section{Problem and simulation model}\label{section2}

\begin{figure}[tb]
	\centering
	\includegraphics*[width=12cm]{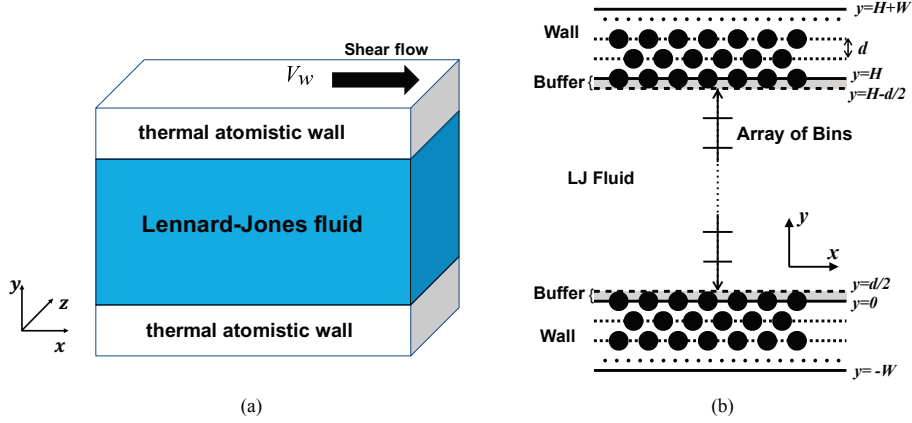}
	\caption{
		A schematic diagram of problem (a) and the setting of the bins for the calculation of local macroscopic quantities (b).
		In figure (a), the LJ fluid is sandwiched between atomistic walls kept at a constant temperature $T^w$.
		Boundary-driven shear flows and force-driven flows are considered.
		The $x$ axis is parallel to the flow direction, and the $y$ axis points in the direction perpendicular to the parallel walls.
		In the $x$ direction, periodic boundary conditions are considered.
		In figure (b), the width of the channel, except for thin layers on the boundaries between the fluid and the channel walls, is uniformly divided into 20 bins, and the local macroscopic quantities are calculated in each bin.
		The thickness of each thin layer is $\frac{d}{2}$, where $d$ is the size of the gap between layers of the FCC lattice structure.
	}\label{FIG.GEOM}
\end{figure}

The LJ fluid between parallel plates, as shown in Fig.~\ref{FIG.GEOM}(a), is considered.
The fluid domain extends $0<y<H$, and the wall domains extend $-W\le y\le 0$ (the lower wall) and $H\le y\le H+W$ (the upper wall).
Both the fluid and the walls are composed of LJ particles that interact with each other via the LJ potential: 
\begin{equation}\label{eq_LJ}
	U(r)=\left\{
		\begin{array}{cc}
			4\varepsilon\left[
				\left(\frac{\sigma}{r}\right)^{12}
			-\left(\frac{\sigma}{r}\right)^6\right],
			& (0<r<r_c),\\
			0,&(r_c\le r).
		\end{array}
		\right.
\end{equation}
Here, $r_c$ is the cut-off parameter, and $\varepsilon$ and $\sigma$ are the units of energy and length of the LJ particles, respectively.

The wall particles are connected to the face-centered cubic (FCC) lattice structure $\{\bm{r}^w_i\}$ by the spring potential, and the temperature of the wall particles is kept at a constant value of $T^w$ by the Langevin thermostat algorithm. 

Thus, the dynamics of the LJ particles are described, for the fluid particles (i.e., ${r_y}_i\in(0,H)$), by
\begin{equation}\label{eq_MD}
	m\ddot{\bm{r}}_i(t)=-\sum_j\frac{\partial U(|\bm{r}_{ij}|)}{\partial \bm{r}_{ij}}
\end{equation}
 and, for the wall particles (i.e., ${r_y}_i \in [-W,0] \cup [H,H+W]$), by
\begin{equation}\label{eq_wall}
m\ddot{\bm{r}}_i(t)=-\sum_j
\frac{\partial U(|\bm{r}_{ij}|)}{\partial \bm{r}_{ij}}
-k_s(\bm{r}_i-\bm{r}^w_i)-\gamma\dot{\bm{r}_i}+\bm{R}(t),
\end{equation}
where $R_\alpha(t)$ ($\alpha=x, y, z$) is the white Gaussian noise that satisfies
\begin{equation}
	<R_\alpha(t)R_\beta(t-s)>=2mk_B T^w\gamma\delta_{\alpha\beta}\delta(s).
\end{equation}
Here, $\bm{r}_i$ represents the position of the $i$th particle, $\bm{r}_{ij}$ is defined as $\bm{r}_{ij}=\bm{r}_i-\bm{r}_j$, $m$ is the mass of an LJ particle, $k_s$ is the spring constant, $\gamma$ is the damping coefficient, and $k_B$ is the Boltzmann constant.
We note that the summation $\sum_j$ applies to both the fluid and wall particles.
We also note that the wall and fluid particles are considered the same in size and mass.

The cut-off length $r_c=2.8$, the spring constant $k_s=10$ and the damping coefficient $\gamma=0.1$ are fixed.
Hereafter, we express quantities in units of mass $m$, energy $\varepsilon$, length $\sigma$, and time $\tau=\sqrt{m\sigma^2/\varepsilon}$.

We note that the temperature of the fluid domain is not artificially controlled by any thermostat algorithm but varies autonomously according to the mass, momentum, and heat transfer between the parallel plates. 
Only the walls are kept at a constant temperature $T^w$ by the Langevin thermostat as Eq.~(\ref{eq_wall}).

The LJ fluid is initially in a uniform liquid state with a density $\rho_0=0.844$ and a temperature $T_0=0.722$, which is near the triple point of the LJ potential.
This initial state of the LJ fluid is produced by a long-time (i.e., 2$\times 10^7$ time steps) quiescent MD simulation of the system shown in Fig.~\ref{FIG.GEOM}(a).
The radial distribution function (RDF) of the initial liquid state is shown in Fig.~\ref{FIG.RDF}(b).

At time $t$=0, the upper wall starts to move from left to right with a speed $V_w$, and the wall-driven shear flow is produced in the fluid domain between the walls.
We note that the temperature of the fluid domain may significantly increase due to the viscous heating when the wall speed is sufficiently large.

\section{Results}\label{sec3}
We consider various channel widths $H$=168, 252, 336, 420, and 504, whereas the side lengths of the simulation box $L_x=L_z=16.8$ and the thickness of the wall $W\simeq 4.1$ are fixed.
The speed of the upper wall also varies as $V_w=$1.0, 1.5, 1.75, 2.0, 2.5, and 3.0 for each channel width $H$.

The MD simulations are performed using the LAMMPS software package \cite{art:95P, LAMMPS}, in which Eq.~\eqref{eq_MD} is time-integrated via the velocity Verlet method with a time-step size of $\Delta t=0.005$.

The width of the channel except the thin buffer layers along the boundaries (see Fig.~\ref{FIG.GEOM}(b)), i.e., $\frac{d}{2}<y<H-\frac{d}{2}$,  is uniformly divided into 20 bins, and in each bin, the local macroscopic quantities, i.e., the density $\rho$, the flow velocity $u_x$, the temperature $T$, and the stress $p_{\alpha\beta}$, are calculated via the following equations.
\begin{equation}\label{eq_rhol}
	\rho(l)=\frac{1}{|V^l_{\rm{bin}}|}\int_{\bm{r}\in V^l_{\rm{bin}}}
	\sum_i\delta(\bm{r}-\bm{r}_i)d\bm{r},
\end{equation}
\begin{equation}\label{eq_uxl}
	\rho(l) u_x(l)=\frac{1}{|V^l_{\rm{bin}}|}\int_{\bm{r}\in V^l_{\rm{bin}}}
	\sum_i\dot r_{x\,i}\,\delta(\bm{r}-\bm{r}_i)d\bm{r},
\end{equation}
\begin{equation}\label{eq_Tl}
	\rho(l) T(l)=\frac{1}{3|V^l_{\rm{bin}}|}\int_{\bm{r}\in V^l_{\rm{bin}}}
	\sum_i\dot ({\bm{r}}_i-u_x(l)\delta_{x\alpha})^2\delta(\bm{r}-\bm{r}_i)d\bm{r},
\end{equation}
\begin{equation}\label{eq_pab}
	p_{\alpha\beta}(l)=\frac{1}{|V^l_{\rm{bin}}|}\int_{\bm{r}\in V^l_{\rm{bin}}}\sum_i
	\left( (\dot r_{\alpha\,i}-u_x(l)\delta_{x\alpha})\dot r_{\beta\,i}+r_{\alpha\,i}f_{\beta\,i}\right)
	\delta(\bm{r}-\bm{r}_i)d\bm{r},
\end{equation}
where the summation $\sum_i$ is taken over all the molecules and $V_{\rm{bin}}^l$ and $|V_{\rm{bin}}^l|$ represent the region of the $l$th bin and the volume of the local bin, respectively.
On the right-hand side of Eq.~(\ref{eq_pab}), $\bm{f}_i$ is the force applied to the $i$th molecule due to the interaction among the ambient molecules, i.e., the right-hand side of Eq.~(\ref{eq_MD}).

The local quantities are also time-averaged in the stationary state after a long time $t_0$ has passed (i.e., $t_0=4\times 10^7\Delta t$), where the instantaneous quantities are sampled every 10 time steps in the interval $t=[t_0,t_0+10^5\Delta t]$ (i.e., $10^4$ samples are averaged for each local quantity).
The standard deviations of the instantaneous local macroscopic quantities shown in Fig.~\ref{FIG.MACRO} are at most 0.023 for the velocity $u_x$, 0.018 for the temperature $T$, 0.0044 for the density $\rho$, and 0.15 for the normal stress $p_{yy}$.

In this section, we mainly show the results for $H$=504. 
The results for other channel widths are given in the supplemental materials.

\subsection{Distribution of macroscopic quantities}
\begin{figure}[tb]
	\centering
	\includegraphics*[width=12cm]{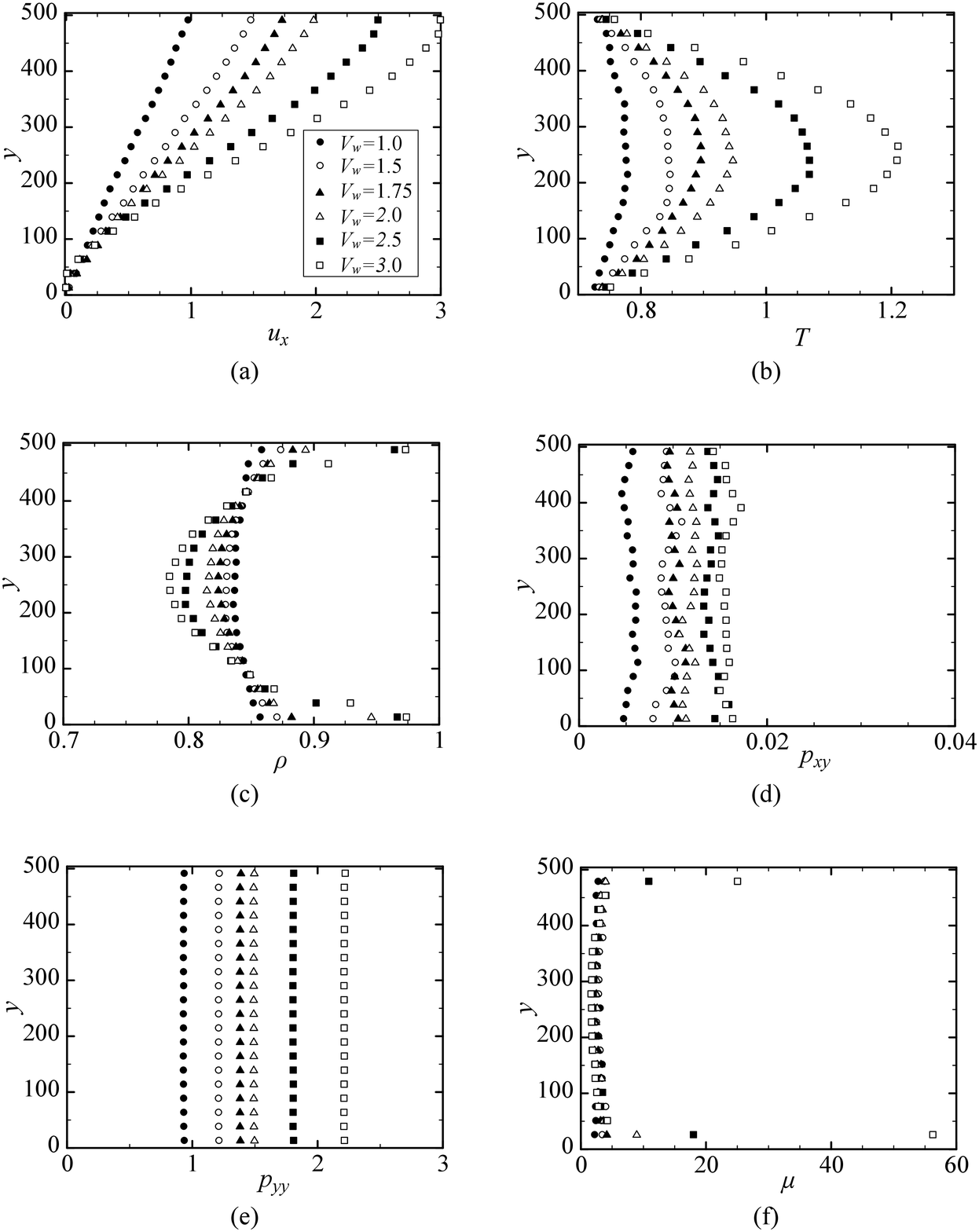}
	\caption{
		The spatial distributions of the macroscopic quantities, i.e., (a) velocity $u_x$, (b) temperature $T$, (c) density $\rho$, (d) shear stress $p_{xy}$, (e) normal stress $p_{yy}$, and (f) local viscosity $\mu$, for different wall velocities $V_w$ in the channel where $H=504$.
	}\label{FIG.MACRO}
\end{figure}
Figure \ref{FIG.MACRO} shows the spatial distributions of the local macroscopic quantities (i.e., velocity, temperature, density, and stress) and local viscosity for different wall speeds $V_w$.
The local viscosity is calculated as $\mu=p_{xy}/(du_x/dy)$.
It is seen that the normal and shear stresses, $p_{yy}$ and $p_{xy}$, are uniformly distributed between the upper and lower walls for all cases.
This fact confirms that the local stresses are balanced so that the flow velocity is in the stationary state.

The other macroscopic quantities spatially vary between the walls.
The temperature increases in the middle region due to viscous heating, while it remains close to the wall temperature near the walls.
By contrast, the local density decreases in the middle but increases near the walls.

Remarkably, for $V_w=$2.5 and 3, we can observe significant jumps in local density and local viscosity near the walls.
Related to the rapid increase in the local viscosity in the vicinity of the wall, the velocity profile becomes nonlinear; i.e., the velocity gradient becomes much smaller near the wall than in the middle of the channel.

In the following text, we put focus on the peculiar behavior observed in the vicinity of the wall when the wall speed is large.

\subsection{Solidification}
\begin{figure}[tb]
	\centering
	\includegraphics*[width=12cm]{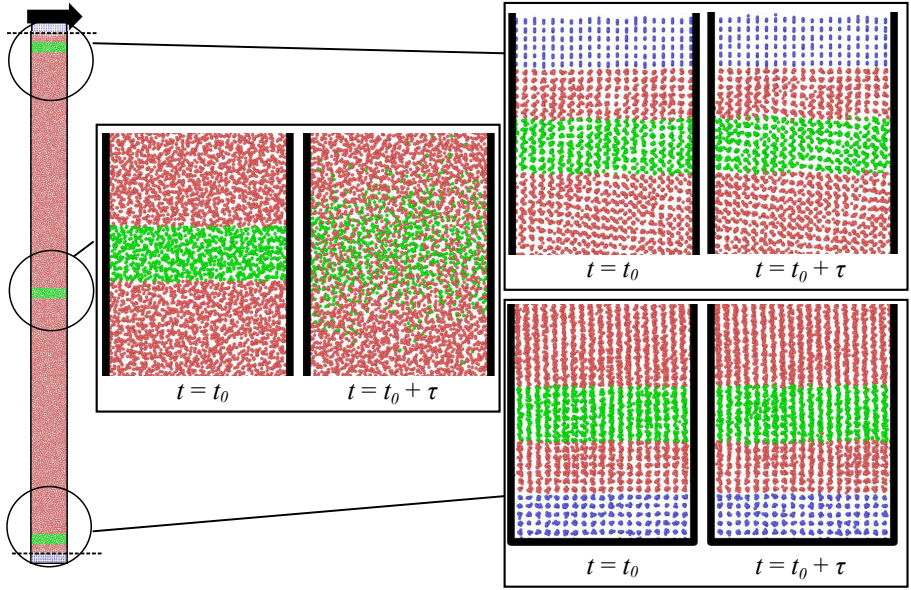}
	\caption{
		Snapshots of molecules in the lower, middle, and upper regions at two different time steps for $H$=504 and $V_w$=3.0.
		The motions of (green-colored) tracer particles after $\tau$=10,000 time steps have passed are observed.
		In the figures, the diameter of each tracer particle is set at 0.3$\sigma$.
	}\label{FIG.TRACE}
\end{figure}
\begin{figure}[tb]
	\centering
	\includegraphics*[width=12cm]{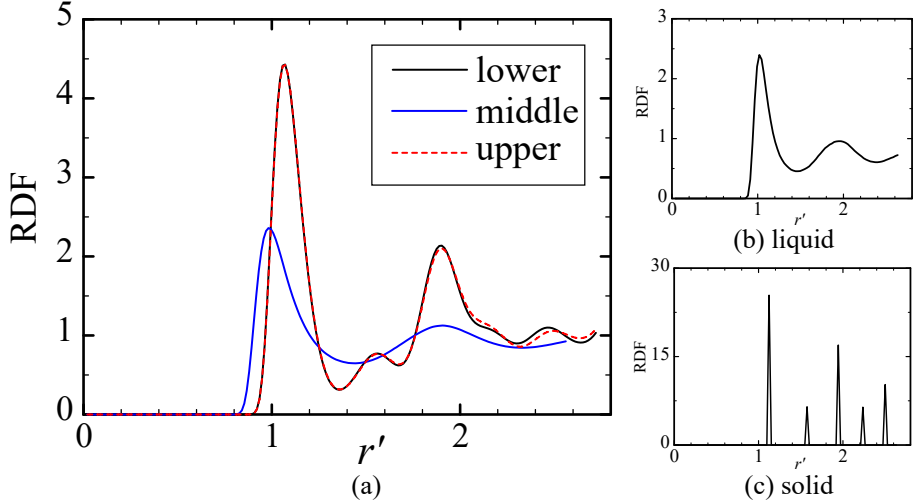}
	\caption{
		Figure (a) shows the RDFs of the molecules in the upper, middle, and lower bins at time $t=t_0$. 
		Figures (b) and (c) show the RDFs for the initial liquid state near the triple point and for the FCC lattice structure of the wall, respectively.
		The horizontal axis is scaled as $r'=r/\rho^{-1/3}$.
		The channel width $H$=504, and the wall speed $V_w$=3.0.
	}\label{FIG.RDF}
\end{figure}

Figure~\ref{FIG.TRACE} shows snapshots of the local distributions of molecules in the lower, middle, and upper regions at two different time steps.
From the motions of tracer particles, we can find the diffusive behaviors of local molecules in the different regions.
It is clearly seen that the tracer molecules in the vicinity of the walls do not diffuse in the lateral direction ($y$-axis) but rather form a crystal-like structure, which is similar to that of the wall molecules.
On the other hand, the molecules in the middle region diffuse in the lateral direction, as is observed in the fluid phase.

{We can also observe the band structure in the solidified layer in Figure~\ref{FIG.TRACE}, where the bands run diagonally right upward near the upper wall.
We note that the band structure can be observed when the wall speed and channel width are sufficiently large, e.g., $V_w\gtrsim 2$ and $H\gtrsim $ (see the supplemental materials).
However, the direction of the band is not always diagonally right upward; instead, it may be the opposite direction or even appear in the $yz$--cross section.
This finding indicates that band formation is not directly related to the flow velocity but rather to the compression of molecules from the bulk region toward the interface of the wall.

It is also seen from the supplemental materials that the solidification of the LJ fluid near the wall occurs  only when the wall speed is sufficiently large, e.g., $V_w\ge 2$.
When the wall speed is small, we observe only a thin absorption layer of molecules on the surface of the wall.
The thickness of the absorption layer is only the length of a few molecules.
}

We also calculate the local RDF in the $l$th bin by
\begin{equation}\label{eq_rdf}
	g^l(r)=\left<\frac{n_i(r)}{4\pi r^2dr\rho_0}\right>_l,
\end{equation}
where $n^i(r)$ counts the number of molecules within the distance [$r$,$r+dr$] from the $i$th molecule in the $l$th bin and $\left<\quad \right>_l$ represents the ensemble average over all the molecules contained in the local bin.
Figure~\ref{FIG.RDF} shows the local RDFs of the molecules in the upper, lower, and middle regions.
It is clearly seen that the RDFs near the walls have similar peak profiles to those of the FCC lattice structure of the wall, while the RDF in the middle region remains in the initial fluid state.
This result also quantitatively confirms that the solidification occurs near the walls and that the lattice structures in the solidified layers are similar to those of the walls.
In Section~\ref{sec_VB}, we will discuss the effect of the wall structure in more detail.

\begin{figure}[htbp]
	\centering
	\includegraphics*[width=8cm]{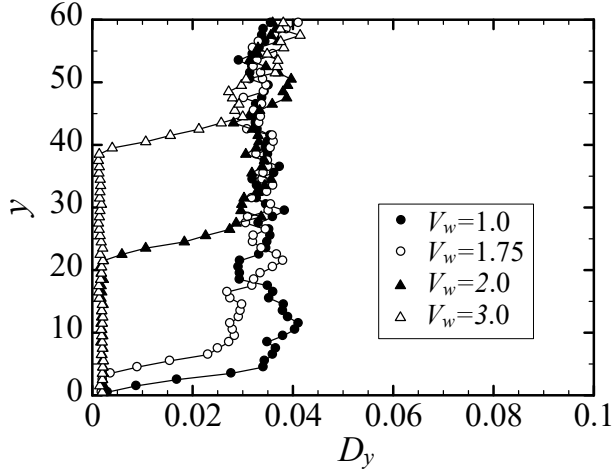}
	\caption{
		The spatial distributions of the local lateral diffusion coefficient $D_y$ defined by Eq.~(\ref{eq_Dy}) for different plate speeds $V_w$=1.0, 1.75, and 2.5 for the channel width $H$=504.
	}\label{FIG.DIFF}
\end{figure}
To distinguish between the fluid and solidified phases, we measure the local lateral diffusion coefficient $D_y$ defined by
\begin{equation}\label{eq_Dy}
D_y=\int_0^\infty \left < \overline{v^i_{y}(t+\tau_0)v^i_y(\tau_0)}\right >_l dt,	
\end{equation}
where $v^i_y$ represents the lateral velocity of the $i$th molecules in the $l$th local bin and $\overline{\cdots}$ represents the time average in $\tau$.

Figure \ref{FIG.DIFF} shows the spatial distributions of the local lateral diffusion coefficient for different wall speeds.
It is clearly seen that the lateral diffusion coefficient $D_y$ is negligibly small in the solidified or absorption layer near the wall.
The solidified layer with a very small lateral diffusion coefficient, e.g., $D_y<0.01$, broadens when the wall speed changes from $V_w=$1.75 to 2.0.

%
\begin{figure}[tb]
	\centering
	\includegraphics[width=10cm]{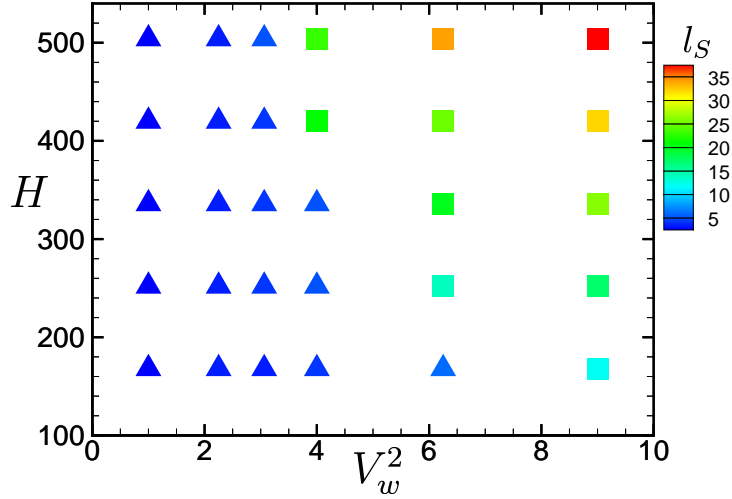}
	\caption{
		Diagram of the phase states near the wall vs. the channel width $H$ and the square of the wall speed $V^2_w$.
		The square symbols show the results obtained when the thickness of the solidified layer $l_S$, which is defined by the thickness of the layer whose lateral diffusion coefficient is as small as $D_y < 0.01$, is $l_S>10$.
		The solid line shows the critical line for the solidification obtained by a crude theoretical estimate, and the dashed line shows the asymptotic limit of the critical line.
			}\label{FIG.DIAGRAM}
\end{figure}
\begin{figure}[tb]
	\centering
	\includegraphics[width=8cm]{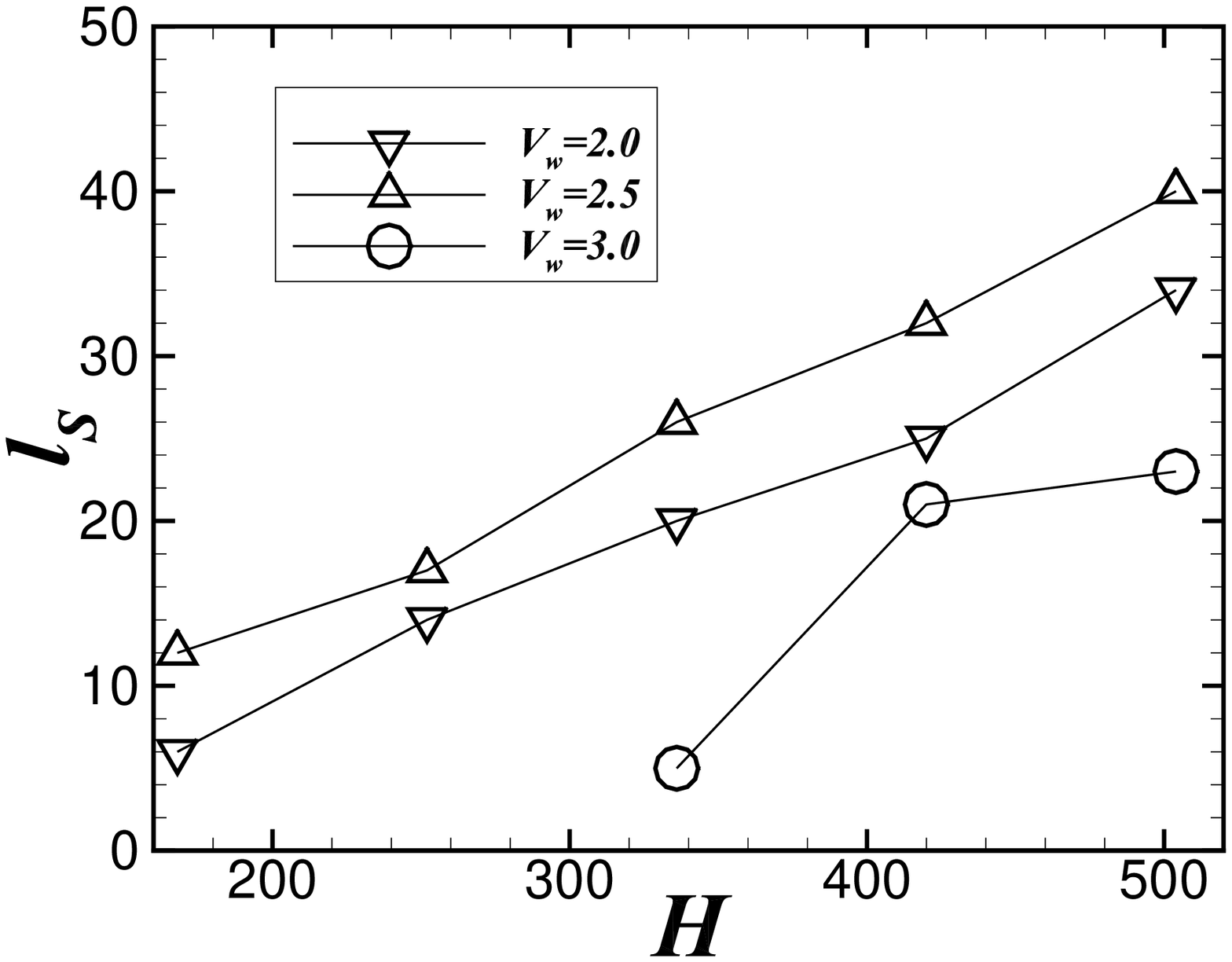}
	\caption{
		The thickness of the solidified layer $l_S$ vs. the channel width $H$ for large wall speeds, i.e., $V_w$=2.0, 2.5, and 3.0.
			}\label{FIG.HLS}
\end{figure}
We summarize the results of the solidification under different parameters in Fig.~\ref{FIG.DIAGRAM}, in which a diagram of the solidification vs. the channel width $H$ and the square of the wall speed $V^2_w$ is shown.
In the figure, the thickness of the solidified layer $l_S$, which is defined by the thickness of the layer where the local lateral diffusion coefficient is smaller than 0.01, i.e., $D_y<0.01$, is indicated by the color legend.
The square symbols $\square$ represent the results obtained when the solidified layer extends far beyond the molecular size, i.e., $l_S>10$.
It is clearly seen that broadened solidification occurs only when the wall speed is sufficiently large, e.g., $V_w \ge 2$.
In the high-speed regime, $V_w\ge 2$, the thickness of the solid layer $l_S$ proportionally increases with the channel width $H$ (see Figure~\ref{FIG.HLS}).
The mechanism underlying these observations will be discussed in the next section.

Incidentally, in Figure~\ref{FIG.DIAGRAM}, the square of the wall speed $V_w^2$ is used as the horizontal axis rather than the wall speed itself.
This is because $V_w^2$ represents the amplitude of the viscous heating relative to the thermal conduction in macroscopic energy transport, i.e.,
\begin{equation}
\frac{\mu V^2_w/H^2}{\lambda \Delta T/H^2}\propto V_w^2,
\end{equation}
where $\lambda$ is the thermal conductivity and $\Delta T$ is a characteristic temperature rise.

\section{Discussion}\label{sec4}

\subsection{Hydrodynamic Explanation}
In this subsection, we consider the mechanism of the solidification from a hydrodynamic point of view.
We suppose a Newtonian fluid with a constant viscosity $\mu$ and the Fourier law of heat conduction with a constant thermal conductivity $\lambda$. 
We also introduce the normalized coordinate $\hat y=y/H$, which is relevant to the hydrodynamic analysis.
In the following part of this section, we consider only the stationary state.
Then, the spatial distribution of temperature is described by
\begin{equation}
-\frac{d^2 T}{d \hat y^2}=\frac{\mu}{\lambda}V_w^2,
\end{equation}
with the boundary condition $T=T_w$ at $\hat y=$0 and 1.
The solution to the above equation is explicitly calculated as
\begin{equation}\label{eq_Ty}
T(\hat y)=4\Delta T \hat y(1-\hat y)+T_w,	
\end{equation}
where $\Delta T=\mu V_w^2/8\lambda$ is the difference in temperatures between the region at the wall and the region in the middle of the channel.

The mass conservation is written as
\begin{equation}\label{eq_massconv}
\int_0^1 \rho(\hat y)d\hat y = \rho_0,
\end{equation}
where $\rho_0$ is the initial density of the LJ fluid.
We can also easily obtain from the momentum balance equation that the bulk pressure is spatially uniform in the stationary state because of the continuity condition $\frac{\partial v_y}{\partial \hat y}$=0.

We suppose that the equation of state $\rho={\cal F}(p,T)$ holds at the local fluid elements in the stationary state even under shear flow and satisfies the conditions $\left(\frac{\partial {\cal F}}{\partial T}\right)_p <0$ and $\left(\frac{\partial {\cal F}}{\partial p}\right)_T >0$.
Then, the bulk pressure $p$ is determined from the equation
\begin{equation}\label{eq_massconv2}
\int_0^1 {\cal F}(p,T(\hat y))d\hat y=\rho_0.	
\end{equation}
This indicates that the bulk pressure does not depend on the channel width $H$ but depends only on the wall speed $V_w$ when the initial states $\rho_0$ and $T_w$ are fixed.
Note that $T(\hat y)$ does not depend on the channel width $H$ in Eq.~(\ref{eq_Ty}).

\begin{figure}[tb]
\includegraphics[width=8cm]{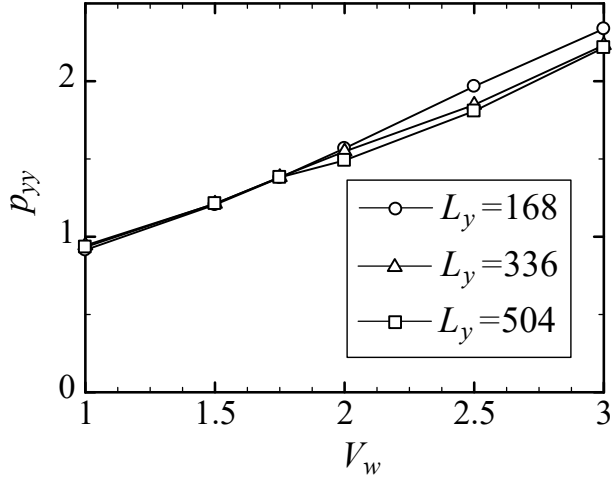}
\caption{The bulk pressure $p_{yy}$ vs. wall speed $V_w$ for various channel widths $H$.}\label{FIG.PvsVw}
\end{figure}

The dependency of the bulk pressure on the wall speed is obtained by taking the derivative of Equation~(\ref{eq_massconv2}) against $V_w$, i.e.,
\begin{equation}\label{eq_dpdv}
\begin{split}
	&\int_0^1 \left(\frac{d p}{d V_w}\right)\left(\frac{\partial {\cal F}}{\partial p}\right)_Td\hat y
	+\int_0^1 \left(\frac{\partial T}{\partial V_w}\right)
	\left(\frac{\partial {\cal F}}{\partial T}\right)_Td\hat y=0,\\
	&\frac{d p}{d V_w}=-\int_0^1 \left(\frac{\partial T}{\partial V_w}\right)
	\left(\frac{\partial {\cal F}}{\partial T}\right)_Td\hat y \Big / \int_0^1 \left(\frac{\partial {\cal F}}{\partial p}\right)_Td\hat y \quad > 0.
	\end{split}	
\end{equation}
Note that from Equation~(\ref{eq_Ty}), $\frac{\partial T}{\partial V_w}>0$ holds at any local position $\hat y\in (0,1)$.
Thus, the bulk pressure dose not depend on the channel width $H$ but monotonically increases with the wall velocity $V_w$.
In fact, in Figure \ref{FIG.PvsVw}, our simulation results demonstrate that the bulk pressure monotonically increases with the wall speed but is less dependent on the channel width.

The local density is described by the equation of state as  $\rho(\hat y)={\cal F}(p,T(\hat y))$, where the bulk pressure $p$ is spatially uniform and the temperature is described by Equation~(\ref{eq_Ty}).
It is seen that the local density monotonically increases while approaching the wall 
(under the condition $\left(\frac{\partial {\cal F}}{\partial T}\right)_p<0$) and takes the maximum value at the wall,
 i.e., $\rho(\hat y) \nearrow \rho_w={\cal F}(p,T_w)$ as $\hat y\rightarrow $ 0 or 1.
This result indicates that the maximum of the local density also monotonically increases with the wall speed $V_w$ but does not depend on the channel width $H$ in the same way as the bulk pressure.

If we suppose that the fluid is solidified when the local density exceeds a critical density $\rho^*$, then the dependency of the maximum density on the wall speed and channel width indicates that the solidification never occurs unless the wall velocity exceeds the critical velocity $V_w^*$, which is obtained from $\rho^*={\cal F}(p(V_w^*),T_w)$, regardless of the channel width $H$.
This concisely explains the observation of the existence of a critical wall speed for solidification in Figure~\ref{FIG.DIAGRAM}.  

The reason why the thickness of the solidified layer, $l_S$, is proportional to the channel width $H$ is also explained; i.e.,
the local density is a monotonic function of $\hat y$ and independent of $H$, so the solidified layer, where the local density is larger than the critical density, i.e., $\rho(\hat y) > \rho^*$, is uniquely determined by the condition $\rho(l_S/H)=\rho^*$ for a given wall velocity.
Thus, the thickness of the solidified layer $l_S$ is proportional to the channel width $H$.

\begin{figure}[tb]
\includegraphics[width=10cm]{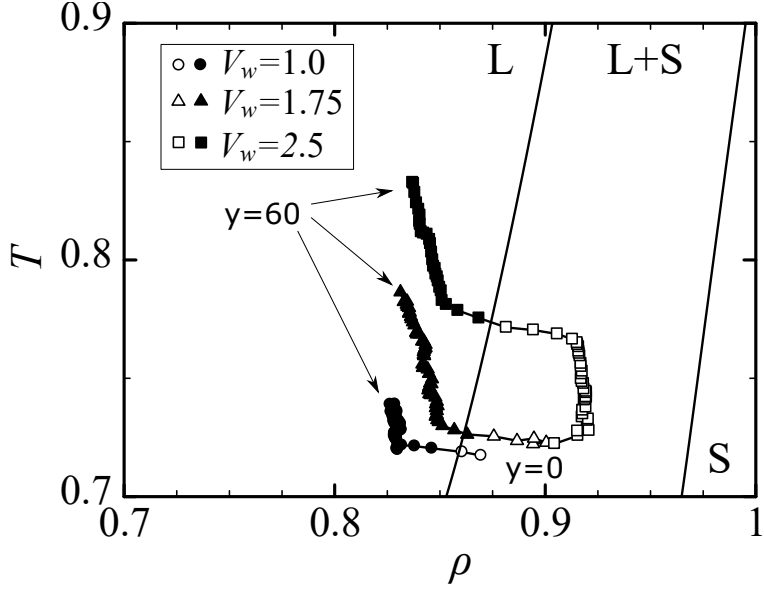}
\caption{The transient behaviors of the local $\rho$--$T$ states between the bulk regime ($y=60$) and the surface of the wall $y=0$ for different wall speeds, i.e., $V_w$=1.0, 1.75, and 2.5.
The initial density $\rho_0$=0.884, the initial temperature $T_w$=0.772, and the channel width $H$=504 are fixed.
In the $\rho$--$T$ plane, the upper left corresponds to the bulk regime, and the lower right corresponds to the vicinity of the wall.
The closed marks indicate that the local lateral diffusion coefficients are smaller than 0.01, i.e., the fluid regime, and the open marks indicate that the local lateral diffusion coefficients are larger than 0.01, i.e., the solid regime.
In the phase diagram, ``L'' represents the liquid phase; ``S'', the solid phase; and ``L+S'', the liquid/solid coexistence phase.
The solidification and melting lines (the left and right solid lines, respectively) are drawn by using the formulas obtained in Ref.~\cite{art:10V}.
}\label{FIG.LOCALPHASE}	
\end{figure}
Figure \ref{FIG.LOCALPHASE} shows the transient behaviors of the local $\rho$--$T$ states between the bulk ($y=60$) and the interface of the wall ($y=0$).
Here, instead of using Eq.~(\ref{eq_rhol}), we calculate the local density by
\begin{equation}
\rho(l)=\frac{1+\int_0^{r_c}4\pi r^2 \rho_0 g^l(r) dr}{\frac{4}{3}\pi r_c^3},
\end{equation}
where the local RDF $g^l(r)$ is defined in Eq.~(\ref{eq_rdf}).
It is seen that the conditions supposed above, i.e., $(\frac{\partial {\cal F}}{\partial T})_p <0$ and $(\frac{\partial {\cal F}}{\partial p})_T>0$, are relevant to the simulation results and, in fact, the maximum density increases with the wall speed $V_w$.

The local density rapidly increases around the solidification line while approaching to the wall from the bulk, and the $\rho$--$T$ state enters into the liquid/solid coexistence regime in the phase diagram.
For a large wall speed $V_w=2.5$, remarkable solidification (or crystallization) is observed even in the liquid/solid coexistence regime, where the thermal expansion becomes very small, i.e, $|\frac{1}{\rho}(\frac{\partial \rho}{\partial T})_p|\ll 1$, as is usually observed in solid materials.
For a small wall speed $V_w<2$, solidification is remarkably not observed, but a thin absorption layer forms on the surface of the wall (see also Fig.~\ref{FIG.DIFF}).

These observations seem to indicate that the LJ molecules in the fluid phase are confined in the vicinity of the wall due to the thermohydrodynamic coupling and that when the density in the confined regime is close to the solidification line, the tightly confined LJ molecules are solidified via the interaction with the wall molecules.
This also indicates that both the wall structure and the $\rho$--$T$ state in the vicinity of the wall strongly affect the solidification near the wall.

\subsection{Effects of the wall structure and the initial state} \label{sec_VB}
\begin{figure}[tb]
	\centering
	\includegraphics[width=12cm]{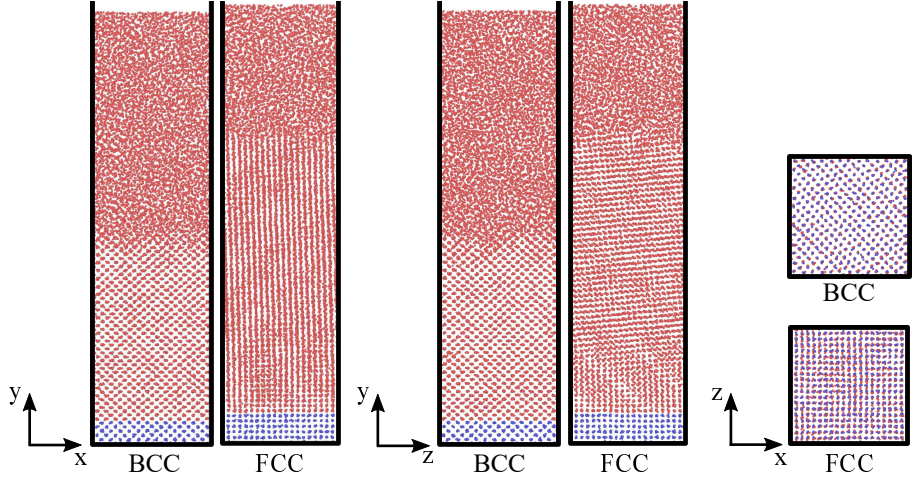}
	\caption{
		Snapshots of the LJ molecules in the vicinity of wall and the wall molecules for the FCC wall (in the left column) and the BCC wall (in the right column).
		The snapshots are shown from three different angles of view.
		In both the FCC and BCC structures, the channel width is $H\simeq 500$, and the wall velocity is $V_w$=3.0 .
	}\label{FIG.FCC_BCC}
\end{figure}

Thus far, we have considered the face-centered-cubic (FCC) lattice structure for the wall molecules and the initial condition of the LJ fluid near the triple point, i.e., $\rho_0$=0.844 and $T_0$=0.722.
In this subsection, we change the wall structure and the initial condition and investigate the effects of the wall structure and the initial condition of the LJ fluid.

Figure \ref{FIG.FCC_BCC} shows snapshots of the molecules composing the bottom wall (i.e., $-H<y<0$) and those in the vicinity of the wall (i.e., {$0<y\leq 60$}) for the channel width $H\simeq 500$ and the wall velocity $V_w$=3.0.
It is seen that solidification occurs with both the FCC and BCC structures.

However, interestingly, the lattice structures of the solidified layer are different from each other.
With the BCC wall, the LJ molecules in the solidified layer also create the BCC lattice structure, although it is known that the FCC structure appears during the crystallization of the LJ molecules in the equilibrium state.
The thickness of the solidified layer is also affected to the wall structure; i.e., the solidified layer for the BCC wall is thinner than that for the FCC wall.

\begin{figure}
	\includegraphics[width=10cm]{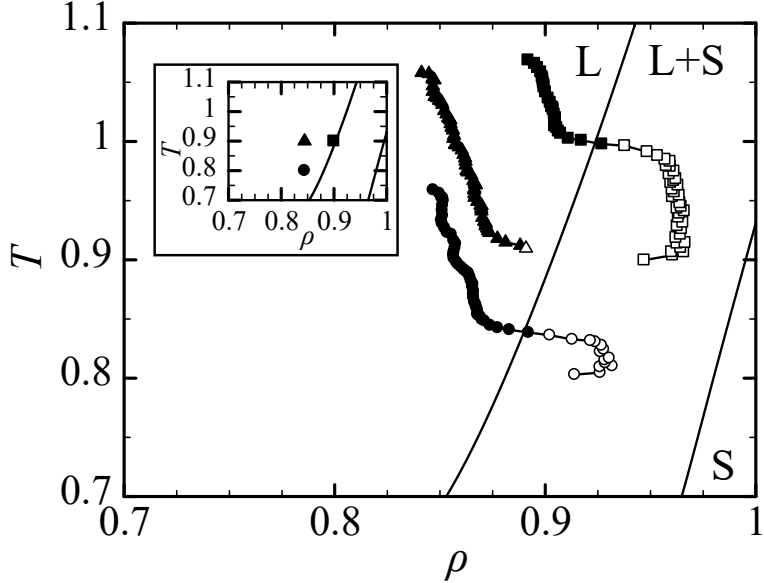}
	\caption{
	The transient behaviors of the local $\rho$--$T$ states between the bulk regime ($y=60$) and the surface of the wall $y=0$ for different initial states, which are shown in the inset, i.e., $\rho_0=0.8442$ and $T=0.8$ for the circle $\bigcirc$, $\rho_0=0.8442$ and $T=0.9$ for the triangle $\bigtriangledown$, and $\rho_0=0.9$ and $T=0.9$ for the square $\Box$. The wall speed $V_w=3.0$ and channel width $H\simeq 500$ are fixed.
	See also the caption in Figure \ref{FIG.LOCALPHASE}.
	}\label{FIG.INIPHASE}
\end{figure}
Figure \ref{FIG.INIPHASE} shows the transient behaviors of local $\rho$--$T$ states from the bulk regime to the surface of the wall for three different initial states of the LJ fluid.
If the initial state is close to the solidification line (i.e., the square $\Box$), remarkable solidification is observed in the liquid/solid coexistence regime in the phase diagram.
Even if the initial state is slightly away from the solidification line (i.e., the circle $\bigcirc$), we can observe that the solidified layer forms in the vicinity of the wall.
However, when the initial state is far from the solidification line (i.e., the triangle $\bigtriangledown$), the local $\rho$--$T$ state cannot approach the solidification line even in the vicinity of the wall, so  solidification does not occur in the vicinity of the wall.

From these observations, we can conclude that the LJ molecules are confined in the vicinity of the wall via thermohydrodynamic transport and that when the local $\rho$--$T$ state is close to the solidification line in the vicinity of the wall, the LJ molecules are solidified due to the interaction with the crystallized wall molecules.

\section{Concluding remarks and perspectives}\label{sec5}
We carried out MD simulations of the thermohydrodynamic lubrication of the LJ fluid between atomistic thermal walls.
A counterintuitive solidification, i.e., {\em viscous heating-induced solidification}, was discovered, in which the LJ fluid is solidified near the wall because of the viscous heating generated in the bulk regime.
It was found that the solidification occurs only when the wall speed is sufficiently large regardless of the channel width, even though the thickness of the solidified layer increases with the channel width.
Band formation was also found in the solidified layer when the channel width is large.

We investigated the solidification mechanism in detail from both macroscopic and microscopic points of views.
It was found that the LJ molecules are confined in the vicinity of the wall via thermohydrodynamic transport and that when the local $\rho$--$T$ state in the vicinity of the wall is close to the solidification line in the phase diagram, the LJ molecules are solidified due to the interaction with the crystallized wall molecules.
The lattice structure of the wall molecules affects that in the solidified layer near the wall; thus, the BCC lattice structure is created in the solidified layer even though the FCC lattice structure is more stable for the LJ molecules.

This study explicitly demonstrates that even for a simple fluid in a simple geometry with smooth boundaries, the thermohydrodynamic coupling in high-speed lubrication flow and the molecular interaction at the interface produce unexpected flow behavior.
In the literature \cite{art:13Aetal}, it is reported that nanoscale surface texture of wall and molecular structure of fluid significantly affect the solidification in confined geometries.
 Investigation on the effects of the roughness of wall surface and the molecular structure of fluid should represent an important future research direction.
 
This study also illustrates that the molecular interaction between wall and fluid significantly affects the phase transition behavior in the confined regime near the wall.
This result gives an important message for the future development of multiscale simulation technologies for large-scale complex flows.

\section*{Acknowledgements} 
This study was financially supported by JSPS KAKENHI Grant Number 16K17554 and 17H01083.

\end{document}